\documentclass[10pt]{bmc_article}
\pdfoutput=1

\usepackage{cite} % Make references as [1-4], not [1,2,3,4]
\usepackage{url} % Formatting web addresses
\usepackage{ifthen} % Conditional
\usepackage{multicol} %Columns
\usepackage{xspace}
\usepackage{hyperref}
\usepackage[utf8]{inputenc} %unicode support
\usepackage[OT1]{fontenc} 

\usepackage{rotating}
\usepackage{colortbl}
\usepackage{color}
\usepackage{comment}

\usepackage{subfigure}

\def\squeezetable{\def\tabular@font{\tiny}}%

\newboolean{publ}

\newenvironment{bmcformat}{\begin{raggedright}\baselineskip20pt\sloppy\setboolean{publ}{false}}{\end{raggedright}\baselineskip20pt\sloppy}

\begin{document}
\begin{bmcformat}

%\title{Non-coding RNAs of bird genomes}
%\title{Wins and losses: Non-coding RNAs of bird genomes}
\title{Conservation and losses of avian non-coding RNA loci}

\author{
Paul P.\ Gardner\correspondingauthor$^{1,2}$
\email{Paul P.\ Gardner\correspondingauthor - paul.gardner@canterbury.ac.nz},
Mario Fasold$^{5,6}$
\email{mario@bioinf.uni-leipzig.de},
Sarah W.\ Burge$^3$
\email{swb@ebi.ac.uk},
Maria Ninova$^4$
\email{Maria.Ninova@postgrad.manchester.ac.uk},
Jana Hertel$^5$
\email{Jana Hertel\correspondingauthor - jana@bioinf.uni-leipzig.de},
Stephanie Kehr$^5$
\email{steffi@bioinf.uni-leipzig.de},
Tammy E.\ Steeves$^1$
\email{tammy.steeves@canterbury.ac.nz},
Sam Griffiths-Jones$^4$
\email{sam.griffiths-jones@manchester.ac.uk}
and
Peter F.\ Stadler\correspondingauthor$^5$
\email{Peter Stadler\correspondingauthor - studla@bioinf.uni-leipzig.de}
}
\address{
\iid(1) School of Biological Sciences, University of Canterbury, Private Bag 4800, Christchurch, New Zealand.
\iid(2) Biomolecular Interaction Centre, University of Canterbury, Private Bag 4800, Christchurch, New Zealand.
\iid(3) European Molecular Biology Laboratory, European Bioinformatics Institute, Hinxton, Cambridge, CB10 1SD, UK.
\iid(4) Faculty of Life Sciences, University of Manchester, Manchester, United Kingdom.
\iid(5) Bioinformatics Group, Department of Computer Science; and Interdisciplinary Center for Bioinformatics, University of Leipzig, H{\"a}rtelstrasse 16-18, D-04107 Leipzig, Germany.
\iid(6) ecSeq Bioinformatics, Brandvorwerkstr.43, D-04275 Leipzig, Germany.
}

\maketitle

\begin{abstract}
Here we present the results of a large-scale bioinformatic annotation
of non-coding RNA loci in 48 avian genomes. Our approach uses
probabilistic models of hand-curated families from the Rfam database
to infer conserved RNA families within each avian genome. We
supplement these annotations with predictions from the tRNA annotation
tool, tRNAscan-SE and microRNAs from miRBase.  We show that a number of
lncRNA-associated loci are conserved between birds and mammals,
including several intriguing cases where the reported mammalian lncRNA
function is not conserved in birds.  We also demonstrate extensive
conservation of classical ncRNAs (e.g., tRNAs) and more recently
discovered ncRNAs (e.g., snoRNAs and miRNAs) in birds. Furthermore, we
describe numerous ``losses'' of several RNA families, and attribute
these to genuine loss, divergence or missing data.  In particular, we
show that many of these losses are due to the challenges associated
with assembling Avian microchromosomes. These combined results illustrate the utility of
applying homology-based methods for annotating novel vertebrate
genomes.
\end{abstract}

\ifthenelse{\boolean{publ}}{\begin{multicols}{2}}{}

\section*{Introduction}

Non-coding RNAs (ncRNAs) are an important class of genes, responsible
for the regulation of many key cellular functions. The major RNA
families include the classical, highly conserved RNAs, sometimes
called ``molecular fossils'', such as the transfer RNAs, ribosomal
RNAs, RNA components of RNase P and the signal recognition particle
\cite{Jeffares:1998}. Other classes appear to have have evolved more
recently, e.g. the small nucleolar RNAs (snoRNAs), microRNAs (miRNAs)
and the long non-coding RNAs (lncRNAs) \cite{Hoeppner:2012}.

The ncRNAs pose serious research challenges, particularly for the
field of genomics. For example, they lack the strong statistical
signals associated with protein coding genes, e.g. open reading
frames, G+C content and codon-usage biases \cite{Rivas:2000}. 

New sequencing technologies have dramatically expanded the rate at
which ncRNAs are discovered and their functions are determined
\cite{cech2014noncoding}. However, in order to determine the full
range of ncRNAs across multiple species we require multiple RNA
fractions (e.g. long and short), in multiple species, in multiple
developmental stages and tissues types. The costs of this approach are
still prohibitive in terms of researcher-time and
finances. Consequently, in this study we concentrate on bioinformatic
approaches, primarily we use homology-based methods (i.e. covariance models
(CMs)). However, we do validate the majority of these predictions
using RNA-seq. The bioinformatic approaches we use remain state of the art for ncRNA bioinformatic
analyses \cite{Sakakibara:1994,Eddy:1994,Nawrocki:2009} and have well
established sensitivity and specificity rates \cite{Freyhult:2007}.
For example, the CM based approach for annotating ncRNAs in genomes requires
reliable alignments and consensus secondary structures of
representative sequences of RNA families, many of which can be found
at Rfam
\cite{Griffiths-Jones:2003,Griffiths-Jones:2005,Gardner:2009,Gardner:2011a,Burge:2013}. These
are used to train probabilistic models that score the likelihood that
a database sequence is generated by the same evolutionary processes as
the training sequences based upon both sequence and structural
information \cite{Sakakibara:1994,Eddy:1994,Nawrocki:2009}.  The
tRNAscan-SE software package uses CMs to accurately predict transfer
RNAs \cite{Lowe:1997,Chan:2009}.

Independent benchmarks of bioinformatic annotation tools have shown
that the CM approaches out-perform alternative methods
\cite{Freyhult:2007}, although their sensitivity can be limited for
rapidly evolving families such as vault RNAs or telomerase RNA
\cite{Menzel:09a}.

%% The functions determined to date for lncRNAs range
%% from regulating chromatin status to chromosomal inactivation
%% \cite{Rinn:2007,Chow:2005}. Yet functional characterisation of these
%% genes is a lengthy and expensive process \cite{Guttman:2009}.

The publication of 48 avian genomes, including the previously
published chicken
\cite{International_Chicken_Genome_Sequencing_Consortium:2004}, zebra
finch \cite{Warren:2010} and turkey \cite{Dalloul:2010} with the
recently published 45 avian genomes
\cite{birds:14,birds:14a,Huang:2013,Zhan:2013,Shapiro:2013,Howard:2013,Li:2014},
provides an exciting opportunity to explore ncRNA conservation in
unprecedented detail.
%{\bf MORE ON BIRD GENOMES. ESP. MACRO \& MICRO
%  CHROMOSOMES.}

In the following we explore the conservation patterns of the major
classes of avian ncRNA loci in further detail.  Using homology search
tools and evolutionary constraints, we have produced a set of genome
annotations for 48 predominantly non-model bird species for ncRNAs
that are conserved across the avian species. This conservative set of
annotations is expected to contain the core avian ncRNA loci. We focus
our report on the unusual results within the avian lineages. These are
either unexpectedly well-conserved ncRNAs or unexpectedly
poorly-conserved ncRNAs. The former are ncRNA loci that were not
expected to be conserved between the birds and the other vertebrates,
particularly those ncRNAs whose function is not conserved in birds.
The latter are apparent losses of ncRNA loci expected to be conserved;
Here, we consider three categories of such ``loss'': First, genuine
gene losses in the avian lineage where ncRNAs well conserved in other
vertebrates are completely absent in birds.  Second, ``divergence''
where ncRNAs have undergone such significant sequence and structural
alternations that homology search tools can no longer detect a
relationship between other vertebrate exemplars and avian varieties.
Third, ``missing'' ncRNAs that failed to be captured in the available,
largely fragmented, avian genomes.  We postulate that the latter
category is likely to be prevalent in comparative avian genome studies
given the distinctive organisation of the avian genome. Namely, the
avian karyotype is characterised by a large number of chromosomes
(average $2n \approx 80$) generally consisting of approximately 5
larger ``macrochromosomes'' and many smaller ``microchromosomes''
\cite{Griffin:2007}. This ’so many, so small’ pattern presents
significant assembly challenges \cite{Ellegren:2005}. Indeed, of the
48 published avian genomes, 20 of which are high-coverage ($>50X$),
only two are chromosomally assembled (chicken and zebra finch;
\cite{birds:14,Warren:2010}).

\section*{Results}

There is substantial gain and loss of lncRNAs and other ncRNA loci
over evolutionary time
\cite{Cabili:2011,Kutter:2012,Hoeppner:2012}. It is difficult to
assess how many of these ``gains'' and ``losses'' are due to limited
bioinformatic sequence alignment tools (these generally fail align
correctly below 60-50\% sequence identity
\cite{Gardner:2005:Nucleic-Acids-Res:15860779}) or due to genuine
gains and losses. Nevertheless, sequence conservation, generally
speaking, provides useful evidence for gene and function conservation.

We have identified 66,879 loci in 48 avian genomes that share sequence
similarity with previously characterised ncRNAs and are conserved in
$>10\%$ of these avian genomes. These loci have been classified into
626 different families, the majority of which correspond to miRNAs and
snoRNAs (summarised in Table~\ref{table:1}). Out of necessity we have
selected the most charismatic families for further discussion. These
include the lncRNAs that appear to be conserved between Mammals and
Aves and the cases of apparent loss of genes that conserved in most
other Vertebrates.

\subsection*{Unusually well conserved RNAs}

The bulk of the ``unusually well conserved RNAs'' belong to the long
non-coding RNA (lncRNA) group.  The lncRNAs are a diverse group of
RNAs that have been implicated in a multitude of functional processes
\cite{Rinn:2007,Chow:2005,Guttman:2009,Ulitsky:2013}. These RNAs have
largely been characterised in mammalian species, particularly human
and mouse. Consequently, we generally do not expect these to be
conserved outside of Mammals. Notable examples include Xist
\cite{Duret:2006} and H19 \cite{Smits:2008}.  There is emerging
evidence for the conservation of ``mammalian'' lncRNAs in Vertebrates
\cite{Chodroff:2010,Ulitsky:2011}), however, like most lncRNAs, the
function of these lncRNAs remains largely unknown. Here, we show the
conservation of several lncRNAs that have been well-characterised in
humans.

The CM based approach is appropriate for most classes of ncRNA, but
the lncRNAs are a particular challenge \cite{Guttman:2009}.  CMs
cannot model the exon-intron structures of spliced lncRNAs, nor do
they deal elegantly with the repeats that many lncRNAs
host. Consequently in the latest release of Rfam the lncRNA families
that were added were composed of local conserved (and possibly
structured elements) within lncRNAs, analogous to the ``domains''
housed within protein sequences \cite{Burge:2013}. Whilst some these
regions may not reflect functional RNA elements but instead regulatory
regions, enhancers or insulators, their syntenic conservation still
provides an indication of lncRNA conservation
\cite{diederichs2014four}.

When analysing the RNA-domain annotations it is striking that many of
the lncRNAs with multiple RNA-domains are consistently preserved in
the birds. The annotations of these domains lie in the same genomic
region, in the same order as in the mammalian homologs. Thus they
support a high degree of evolutionary conservation for the entire
lncRNA. In particular the HOXA11-AS1, PART1, PCA3, RMST, Six3os1,
SOX2OT and ST7-OT3 lncRNAs have multiple, well conserved RNA-domains
(See Figure~\ref{fig:1}). The syntenic ordering of these seven lncRNAs
and the flanking genes are also preserved between the human and
chicken genomes (data not shown). We illustrate this in detail for the
HOTAIRM1 lncRNA (See Supplemental Figures~13\&14).

The conservation of these ``human'' lncRNAs among birds suggests they
may also be functional in birds. But what these functions may be is not
immediately obvious. For example, PART1 and PCA3 are both described as
prostate-specific lncRNAs that play a role in the human
androgen-receptor pathway
\cite{Bussemakers:1999,Lin:2000,Ferreira:2012}. Birds lack a prostate
but both males and females express the androgen receptor (AR or NR3C4)
in gonadal and non- gonadal tissue
\cite{Yoshimura:1993,Veney:2004,Fuxjager:2012,Leska:2012}. Thus, we
postulate that PART1 and PCA3 also play a role in the
androgen-receptor pathway in birds but whether the expression of these
lncRNAs are tissue specific is unknown at present.

The HOX cluster lncRNAs HOTAIRM1 (5 RNA-domains), HOXA11-AS1 (6
RNA-domains), and HOTTIP (4 RNA domains) are conserved across the
Mammalian and Avian lineages. In the human genome they are located in
the HOXA cluster (hg coordinates chr7:27135743-27245922), one of the
most highly conserved regions in vertebrate genomes
\cite{PascualAnaya:13}, in antisense orientation between HoxA1 and
HoxA2, between HoxA11 and HoxA13, and upstream of HoxA13,
respectively. Conservation and expression of HOTAIRM1 and HOXA11-AS1
within the HOXA cluster has been studied in some detail in marsupials
\cite{Yu:12}.  Of the 15 RNA-domains five and six representing all
three lncRNAs were recovered in the alligator and turtle genomes. All
of them appear in the correct order at the expected, syntenically
conserved positions within the HOXA cluster.  In the birds, where two
or more of the HOX cluster lncRNA RNA-domains were predicted on the
same scaffold, this gene order and location within HOX was also
preserved.

%egrep 'HOXA11-AS1|HOTAIR|HOTTIP' clans_competed/*gff | perl -lane 'if(/\/(\S+?)\-.*gff:(\S+).*\-id=(\S+);eval/ or /\/(\S+?)\-.*gff:(\S+).*Alias=(\S+);Not/){print "$1\t$2\t$3\t$F[6]"}'
%egrep 'PCA3' clans_competed/*gff | perl -lane 'if(/\/(\S+?)\-.*gff:(\S+).*\-id=(\S+);eval/ or /\/(\S+?)\-.*gff:(\S+).*Alias=(\S+);Not/){print "$1\t$2\t$3\t$F[6]"}'
%egrep 'RMST' clans_competed/*gff | perl -lane 'if(/\/(\S+?)\-.*gff:(\S+).*\-id=(\S+);eval/ or /\/(\S+?)\-.*gff:(\S+).*Alias=(\S+);Not/){print "$1\t$2\t$3\t$F[6]"}'

The majority ($>80\%$) of genome-wide association studies of cancer
identify loci outside of protein-coding genes
\cite{Cheetham:2013}. Many of these are loci are now known to be
transcribed into lncRNAs. Furthermore, many lncRNAs are differentially
expressed in tumorous tissues \cite{Chan:2002}, suggesting further
mechanistic links with the aberrant gene expression associated cancer
progression. In our work we have identified three examples of these
that are also conserved in the birds are described below.

The RMST (Rhabdomyosarcoma 2 associated transcript) RNA-domains 6, 7,
8, and 9 are conserved across the birds. In each bird the gene order
was also consistent with the human ordering. In the alligator and
turtle an additional RNA-domain was predicted in each, these were
RNA-domains 2 and 4 respectively, again the ordering of the domains
was consistent with human. This suggests that the RMST lncRNA is
highly conserved. However, little is known about the function of this
RNA. It was originally identified in a screen for differentially
expressed genes in two Rhabdomyosarcoma tumor types \cite{Chan:2002}.

In addition, the lncRNA DLEU2 is well conserved across the
vertebrates, it is a host gene for two miRNA genes, miR-15 and miR-16,
both of which are also well conserved across the vertebrates (See
Supplemental Figure~2). DLEU2 is thought to be a tumor-suppressor gene
as it is frequently deleted in malignant tumours
\cite{Lerner:2009,Klein:2010}.

The NBR2 lncRNA and BRCA1 gene share a bidirectional promotor
\cite{Xu:1997}. Both are expressed in a broad range of
tissues. Extensive research on BRCA1 has shown that it is involved in
DNA repair \cite{Moynahan:1999}. The function of NBR2 remains unknown,
yet its conservation across the vertebrates certainly implies a
function (See Figure~\ref{fig:1}). We note that the function for this
locus may be at the DNA level, however, function at the RNA level
cannot be ruled out at this stage.

Of the other classes of RNAs, none showed an unexpected degree of
conservation or expansion within the avian lineage. The only exception
being the snoRNA, SNORD93. SNORD93 has 92 copies in the tinamou
genome, whereas it only has 1-2 copies in all the other vertebrate
genomes.

\subsection*{Unexpectedly poorly conserved  ncRNAs: genuine loss, divergence or missing  data?}

\subsubsection*{Genuine loss}

The overall reduction in avian genomic size has been extensively
discussed elsewhere \cite{Organ:2007}. Unsurprisingly, this reduction
is reflected in the copy-number of ncRNA genes. Some of the most
dramatic examples are the transfer RNAs and pseudogenes which average
$\sim900$ and $\sim580$ copies in the human, turtle and alligator
genomes, the average copy-numbers of these drop to $\sim280$ and
$\sim100$ copies in the avian genomes. In addition to reduction in
copy-number, the absence of several, otherwise ubiquitous vertebrate
ncRNAs, in the avian lineage are suggestive of genuine gene loss.

Namely, mammalian and amphibian genomes contain three loci of
clustered microRNAs from the mir-17 and mir-92 families
\cite{Tanzer:04}. One of these clusters (cluster II, with families
mir-106b, mir-93 and mir-25) was not found in turtles, crocodiles and
birds (see Supplemental Figure~6). In addition, the microRNA family
let-7 is the most diverse microRNA family with 14 paralogs in
human. These genes also localize in 7 genomic clusters, together with
mir-100 and mir-125 miRNA families (see previous study on the
evolution of the let-7 miRNA cluster in \cite{Hertel:2012}). In Sauropsids we
observed that cluster A - which is strongly conserved in vertebrates
has been completely lost in the avian lineage.  Another obvious loss
in birds is cluster F, containing two let-7 microRNA paralogs. Cluster
H, on the other hand has been retained in all oviparous animals and
completely lost later, after the split of Theria (see Supplemental
Figure~7).

\subsubsection*{Divergence}

In order to determine to what extent the absence of some ncRNAs from
the infernal-based annotation is caused by sequence divergence beyond
the thresholds of the Rfam CMs, we complemented our analysis by
dedicated searches for a few of these RNA groups. Our ability to find
additional homologs for several RNA families that fill gaps in the
abundance matrices (Figure~1) strongly suggests that conspicuous
absences, in particular of LUCA and LECA RNAs, are caused by
incomplete data in the current assemblies and sequence divergence
rather then genuine losses.

Vertebrate Y RNAs typically form a cluster comprising four
well-defined paralog groups Y1, Y3, Y4, and Y5. In line with
\cite{Mosig:07a} we find that the Y5 paralog family is absent from all
bird genomes, while it is still present in both alligator and turtle,
see Supplemental Figure~4. Within the avian lineage, we find a
conserved Y4-Y3-Y1 cluster. Apparently, broken-up clusters are in most
cases consistent with breaks (e.g. ends of contigs) in the available
sequence assemblies.  In several genomes we observe one or a few
additional Y RNA homologs unlinked to the canonical Y RNA
cluster. These sequences can be identified unambiguously as derived
members of one of the three ancestral paralog groups, they almost
always fit less well to the consensus (as measured by the CM bit score
of paralog group specific covariance models) than the paralog linked
to cluster, and there is no indication that any of these additional
copies is evolutionarily conserved over longer time scales. We
therefore suggest that most or all of these interspersed copies are in
fact pseudogenes (see below).

\subsubsection*{Missing data}

Seven families of ncRNAs were found in some avian genomes but not
others (Figure~\ref{fig:1}). These families range in conservation level
from being ubiquitous to cellular-life (RNase P and tRNA-sec), present
in most Bilateria (vault), present in the majority of eukaryotes
(RNase MRP, U4atac and U11) and present in all vertebrates
(telomerase) \cite{Hoeppner:2012}. Therefore, the genuine loss or even
diversification of these ncRNA families in the avian lineage is
unlikely. Rather, this lack of phylogenetic signal, combined with the
fragmented nature of the vast majority of these genomes described
above (i.e., of the 48 avian genomes, only the chicken and zebra finch
are chromosomally assembled \cite{birds:14,Warren:2010}), suggests the most likely
explanation is that these ncRNA families are indicative of missing
data. Indeed, of the seven missing ncRNA families, six where found in
the chicken genome and three were found in the zebra finch
genome. Furthermore, only one of these (RNase MRP) is found on a
macrochromosome, and all remaining missing ncRNAs are found on
microchromosomes (see Supplemental Table~1). A Fisher’s exact test
showed that there is significantly more missing ncRNAs on
microchromosomes than macrochromosomes ($P<10^{16}$ for both the
chicken and zebra finch). Thus, we suggest that many of
these ncRNAs families are missing because: (1) they are predominantly
found on microchromosomes [this study] and (2) the vast majority of
avian microchromosomes remain unassembled \cite{birds:14,Ellegren:2005}.

To wit, we performed dedicated searches for a selection of these
missing ncRNA families. Here, tRNAscan is tuned for specificity and
thus misses several occurrences of tRNA-sec that are easily found in
the majority of genomes by \texttt{blastn} with $E\le 10^{-30}$. In some cases the
sequences appear degraded at the ends, which is likely due to low
sequence quality at the very ends of contigs or scaffolds. A \texttt{blastn}
search also readily retrieves additional RNase P and RNAse MRP RNAs in
the majority of genomes, albeit only the best conserved regions are
captured. In many cases these additional candidates are incomplete or
contain undetermined sequence, which explains why they are missed by
the CMs \cite{Stadler:09b,Kolbe:2009}.

\section*{Pseudogenes}

%{\bf Briefly mention the reduction in number of pseudogenes. Pick a
%  few key human ones and compare with the birds. }

Non-coding RNA derived pseudogenes are a major problem for many ncRNA
annotation projects. The human genome, for example, contains $>1$
million Alu repeats, which are derived from the SRP RNA
\cite{Wheeler:2013}. The existing Rfam annotation of the human genome,
in particular, contains a number of problematic families that appear
to have been excessively pseudogenised. The U6 snRNA, SRP RNA and Y
RNA families have 1,371, 941 and 892 annotations in the human
genome. These are a heterogenous mix of pseudogenised, paralogous,
diverged or functional copies of these families. Unfortunately, a
generalised model of RNA pseudogenes has not been incorporated into
the main covariance model package, Infernal. An approach used by tRNAscan
\cite{Lowe:1997}, is, in theory, generalisable to other RNA families
but this remains a work in progress. 

It is possible that the avian annotations also contains excessive
pseudogenes. However, it has previously been noted that avian genomes
are significantly smaller than other vertebrate species
\cite{International_Chicken_Genome_Sequencing_Consortium:2004}. We
have also noted a corresponding reduction in the number of paralogs
and presumed ncRNA-derived pseudogenes in the avian genomes (See
Supplemental Figure~12). The problematic human families, U6 snRNA, SRP
RNA and Y RNA have, for example, just 26, 4 and 3 annotations
respectively in the chicken genome and 13, 3 and 3 annotations
respectively, on average, in the 48 avian genomes used here.
Therefore, we conclude that the majority of our annotations are in fact
functional orthologs.

\section*{Experimentally confirmed ncRNAs}

The ncRNAs presented here have been identified using homology models
and are evolutionarily conserved in multiple avian species. In order
to further validate these predictions we have used strand-specific
total RNA-seq and small RNA-seq of multiple chicken tissues. After
mapping the RNA-seq data to the chicken genome (see Methods for
details), we identified a threshold for calling a gene as expressed by
limiting our estimated false-positive rate to approximately 10\%. This
FDR was estimated using a negative control of randomly selected,
un-annotated regions of the genome. Since some regions may be
genuinely expressed, the true FDR is potentially lower than
10\%. Overall, the number of ncRNAs we have identified in this work
that are expressed above background levels is 865 (72.4\%) (see
Table~\ref{table:1}). This shows that 7.0 times more of our ncRNAs are
expressed than expected by chance (Fisher's exact test:
$P<10^{16}$). This number is an underestimate of the fraction of our
annotations that are genuinely expressed, as only a fraction of the
developmental stages and tissues of chicken have been characterized
with RNA-seq. Furthermore, some ncRNAs are expressed in highly
specific conditions \cite{mercer2008specific,johnston2003microrna}.

The classes of RNAs where the majority of our annotations were
experimentally confirmed includes microRNAs, snoRNAs, cis-regulatory
elements, tRNAs, SRP RNA and RNase P/MRP RNA. The RNA-seq data could
not provide evidence for a telomerase RNA transcript, which are only
generally only expressed in embryonic, stem or cancerous tissues. Only
a small fraction of the 7SK RNA, the minor spliceosomal RNAs and the
lncRNAs could be confirmed with the 10\% FDR threshold. There are a
number of possible explanations for this: the multiple copies of the
7SK RNA may be functionally redundant and can therefore compensate for
one another; The minor spliceosome is, as the name suggests, a rarely
used alternative spliceosome; and the lncRNAs are generally expressed
at low levels under specific conditions
\cite{mercer2008specific,mercer2012targeted}.

%totalExpressed[817] totalNegControlsExpressed[117] totalGenes[1194]

\section*{Conclusions}

%Talking points:
%1. several classical RNAs (e.g. spliceosomal, telomerase, RNase MRP
%RNAs) have diversified in birds relative to other Eukaryotes.
%2. several lncRNA families are conserved between mammals and
%birds. Several of these are characterised by expression in tissues
%that are not present in birds (e.g. the prostate).
%3. several regulatory RNAs, snoRNAs and microRNAs, have been lost
%since the birds diverged from the other vertebrates. Indicating many
%avian-specific alterations in gene regulation.

In this work we have provided a comprehensive annotation of non-coding
RNAs in genome sequences using homology-based methods. The
homology-based tools have distinct advantages over experimental-based
approaches as not all RNAs are expressed in any particular tissue-type
or developmental-stage, in fact some RNAs have extremely specific
expression profiles, e.g. the lsy-6 microRNA \cite{Johnston:2003}.  We
have identified previously unrecognised conservation of ncRNAs in
avian genomes and some surprising ``losses'' of otherwise well
conserved ncRNAs. We have shown that most of these losses are due to
difficulties assembling avian microchromosomes rather than \emph{bona
  fide} gene loss. A large fraction of our annotations have been
confirmed using RNA-seq data, which also showed a 7-fold enrichment of
expression within our annotations relative to unannotated regions.

The collection of ncRNA sequences is generally biased towards model
organisms \cite{Gardner:2010,Hoeppner:2012}. However, we have shown
that using data from well studied lineages such as mammals can also
result in quality annotations of sister taxa such as Aves.

In summary, these results indicate we are in the very early phases of
determining the functions of many RNA families. This is illustrated by
the fact that the reported functions of some ncRNAs are
mammal-specific, yet these are also found in bird genomes.

\section*{Methods}

The 48 bird genome sequences used for the following analyses are available from
the phylogenomics analysis of birds website \cite{birdphylogen}.

Bird genomes were searched using the cmsearch program from INFERNAL
1.1 and the covariance models from the Rfam database
v11.0 \cite{Gardner:2011a,Burge:2013}. All matches above the curated GA
threshold were included. Subsequently, all hits with an E-value
greater than 0.0005 were discarded, so only matches which passed the
model-specific GA threshold and had an E-value smaller than 0.0005
were retained. The Rfam database classifies non-coding RNAs into
hierarchical groupings. The basic units are ``families'' which are
groups of homologous, alignable sequences; ``clans'' which are groups
of un-alignable (or functionally distinct), homologous families; and
``classes'' which are groups of clans and families with related
biological functions e.g. spliceosomal RNAs, miRNAs and snoRNAs
\cite{Griffiths-Jones:2003,Griffiths-Jones:2005,Gardner:2009,Gardner:2011a,Burge:2013};
these categories have been used to classify our results.

In order to obtain good annotations of tRNA genes we ran the
specialist tRNA-scan version 1.3.1 annotation tool. This method also
uses covariance models to identify tRNAs. However it also uses some
heuristics to increase the search-speed, annotates the Isoacceptor
Type of each prediction and uses sequence analysis to infer if
predictions are likely to be functional or tRNA-derived pseudogenes
\cite{Lowe:1997,Chan:2009}.

Rfam matches and the tRNA-scan results for families belonging to the
same clan were then ``competed'' so that only the best match was
retained for any genomic region \cite{Gardner:2011a}.  To further
increase the specificity of our annotations we filtered out families
that were identified in four or fewer of the 51 vertebrate species we
have analysed in this work. These filtered families largely
corresponded to bacterial contamination within the genomic sequences.

999 microRNA sequence families, previously annotated in at least one
vertebrate, were retrieved from miRBase (v19). Individual sequences or
multiple sequence alignments were used to build covariance models with
INFERNAL (v1.1rc3), and these models were searched against the 48 bird
genomes, and the genomes of the American alligator and the green
turtle as outgroups. Hits with e-value $<10$ realigned with the query
sequences and the resultant multiple sequence alignments manually
inspected and edited using RALEE.

An additional snoRNA homology search was performed with snoStrip
\cite{Bartschat:2013}. As initial queries we used deutorostomian
snoRNA families from human \cite{Lestrade:2006}, platypus
\cite{Schmitz:2008}, and chicken \cite{Shao:2009}.

%SnoRNA annotations overlapping miRNAs annotations where manually
%inspected \ppg{and?...}.

The diverse sets of genome annotations were combined and filtered,
ensuring conservation in 10\% or more of the avian genomes. We
collapsed the remaining overlapping annotations into a single
annotation. We also generated heatmaps for different groups of ncRNA
genes (see Figure~\ref{fig:1} and Supplemental Figure~1-3). All the
scripts and annotations presented here are available from Github
\cite{gitrepo}.

Chicken ncRNA predictions were validated using two separate RNA-seq
data sets. The first data set (Bioproject PRJNA204941) contains 971
million reads and comprises 27 samples from 14 different chicken
tissues sequenced on Illumina HiSeq2000 using a small RNA-seq
protocol. The second data set (SRA accession SRP041863) contains 1,46
billion Illumina HiSeq reads sequenced from whole chicken embryo RNA
from 7 stages using a strand-specific dUTP protocol. Raw reads were
checked for quality and adapters clipped if required by the
protocol. Preprocessed reads were mapped to the galGal4 reference
genome using SEGEMEHL short read aligner \cite{segemehl} and then
overlapped with the ncRNA annotations.

\section*{Acknowledgements}

Erich Jarvis (Duke University), Guojie Zhang (BGI-Shenzhen \&
University of Copenhagen) and Tom Gilbert (University of Copenhagen)
for access to data and for invaluable feedback on the manuscript.

Magnus Alm Rosenblad (Univ. of Gothenburg) and Eric Nawrocki (HHMI
Janelia Farm) for useful discussions. Matthew Walters for assistance
with figures.

We thank Fiona McCarthy (University of Arizona) and Carl
Schmidt (University of Delaware) as well as Matt Schwartz (Harvard)
and Igor Ulitsky (Weizmann Institute of Science) for providing the
RNA-seq data as part of the Avian RNAseq consortium.

Thanks to @ewanbirney for the following timely tweet: ``So ... missing
orthologs to chicken often mean 'gene might be on the
microchromosome''.

We thank the anonymous referees for providing invaluable suggestions
that improved this work. 

%%%%%%%%%%%%%%%%%%%%%%
%% The Bibliography %%
%%
%% ---------------------------------
%% BioMedCentral bibtex .BST file will be used to
%% create a .BBL file which includes the BMC XML.
%% Note that the displayed Bibliography will not be
%% exactly as specified in the online Instructions for Authors

{\ifthenelse{\boolean{publ}}{\footnotesize}{\small}
\bibliographystyle{bmc_article} % Style BST file
 \bibliography{bird} } % Bibliography file (usually '*.bib' )

%%%%%%%%%%%

\clearpage
\newpage

\section*{Figures}
  \subsection*{Figure 1 - Heatmaps}

Heatmaps showing the presence/absence and approximate genomic
copy-number of ``unusually, well conserved RNAs'' (particularly the
lncRNAs) on the left and families that have been identified as RNA
losses, divergence or missing data. In several cases functionally
related families have also been included, e.g. the RNA components of
the major and minor spliceosomes: U1, U2, U4, U5 and U6; and U11, U12,
U4atac, U5 and U6atac, respectively.

 \begin{figure}[ht]
   \centering
   \includegraphics[width=0.95\textwidth]{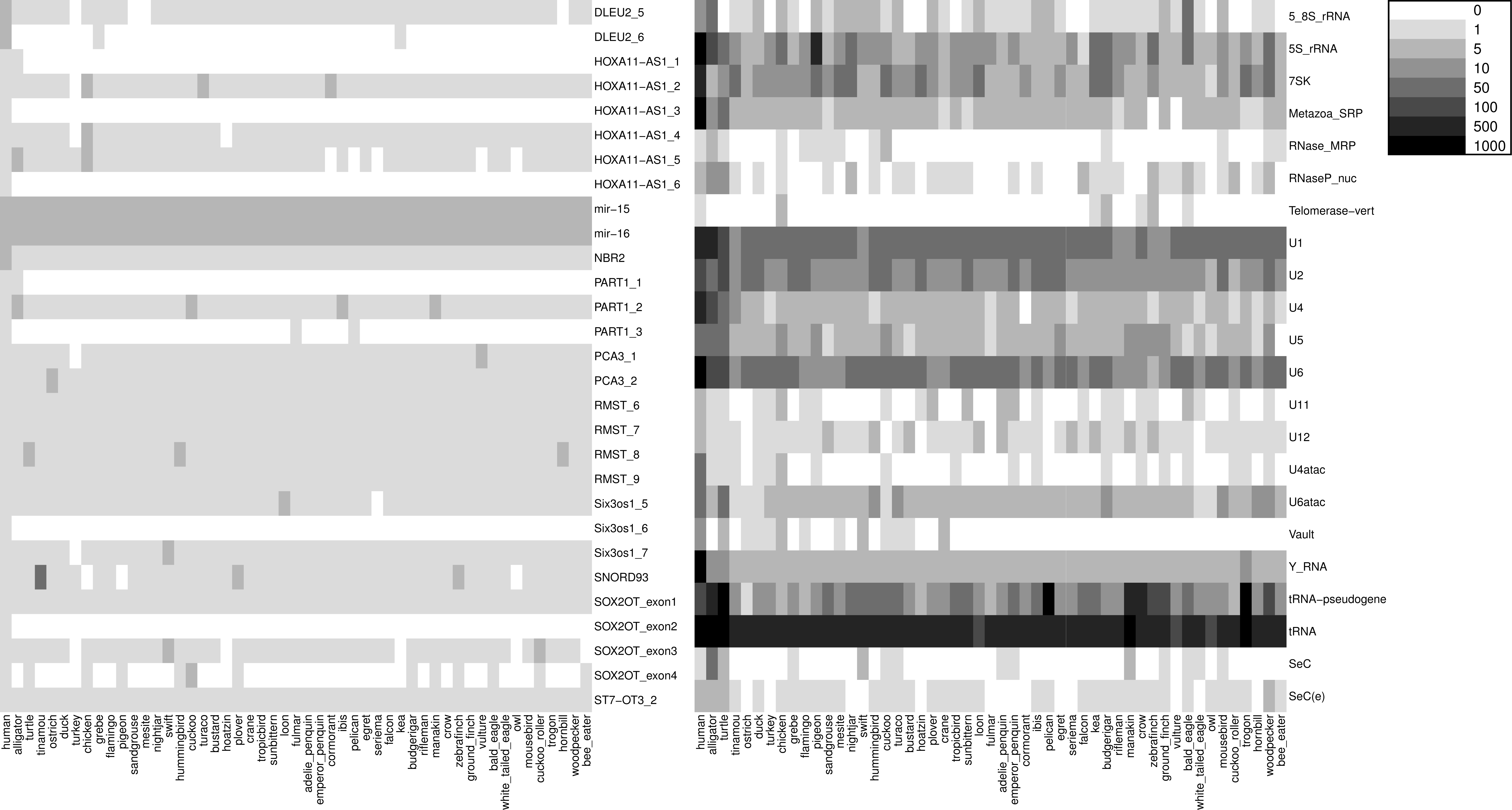}
   \caption{}
   \label{fig:1}
 \end{figure}

\clearpage
\newpage

\subsection*{Table 1 - A summary of ncRNA genes in human, chicken and all bird genomes}

This table contains the total number of annotated ncRNAs from
different RNA types in human, the median number for each of the 48
birds and chicken. The number of chicken ncRNA that show evidence for
expression is also indicated (the percentage is given in parentheses).
The threshold for determining expression was selected based upon a
false positive rate of less than 10\%.

%% \begin{tabular}{|r|r|r|r|l|}
%% \hline 
%% \multicolumn{5}{|l|}{{\bf ncRNA genes in human, chicken and all bird genomes}}\\
%% \hline 
%% Number in human & median(48 birds) & Number in chicken & Chicken ncRNAs         & RNA type\\
%%                 &                  &                   & confirmed with RNA-seq & \\
%% \hline
%% 62&25.0&34&12 (35.3\%) &Long non-coding RNA\\ 
%% 356&499.5&427&272 (63.7\%) &microRNA\\ 
%% 281&120.0&106&88 (83.0\%) &C/D box snoRNA\\ 
%% 336&85.5&68&46 (67.6\%) &H/ACA box snoRNA\\ 
%% 34&13.0&12&11 (91.7\%) &Small cajal body RNA\\ 
%% 1754&48.5&71&29 (40.8\%) &Major spliceosomal RNA\\ 
%% 58&3.0&6&2 (33.3\%) &Minor spliceosomal RNA\\ 
%% 525&82.0&122&79 (64.8\%) &Cis-regulatory element\\ 
%% 316&6.5&9&3 (33.3\%) &7SK RNA\\ 
%% 1&0.0&2&0 (0.0\%) &Telomerase RNA\\ 
%% 9&0.0&2&1 (50.0\%) &Vault RNA\\ 
%% 892&3.0&3&2 (66.7\%) &Y RNA\\ 
%% 1084&173.5&300&256 (85.3\%) &Transfer RNA\\ 
%% 80&9.5&4&2 (50.0\%) &Transfer RNA pseudogene\\ 
%% 941&3.0&4&2 (50.0\%) &SRP RNA\\ 
%% 607&7.0&22&10 (45.5\%) &Ribosomal RNA\\ 
%% 4&1.0&2&2 (100.0\%) &RNase P/MRP RNA\\ 
%% \hline
%% 7340&1080.0&1194&817 (68.4\%) &Total\\ 
%% \hline
%% \end{tabular}

\begin{table}
\begin{tabular}{|r|r|r|r|l|}
\hline 
\multicolumn{5}{|l|}{{\bf ncRNA genes in human, chicken and all bird genomes}}\\
\hline 
                &                  &                   & Chicken ncRNAs & \\
Number in human & median(48 birds) & Number in chicken & confirmed with RNA-seq & RNA type\\
\hline
62&25.0&34&12 (35.3\%) &Long non-coding RNA\\ 
356&499.5&427&280 (65.6\%) &microRNA\\ 
281&120.0&106&90 (84.9\%) &C/D box snoRNA\\ 
336&85.5&68&48 (70.6\%) &H/ACA box snoRNA\\ 
34&13.0&12&12 (100.0\%) &Small cajal body RNA\\ 
1754&48.5&71&32 (45.1\%) &Major spliceosomal RNA\\ 
58&3.0&6&3 (50.0\%) &Minor spliceosomal RNA\\ 
525&82.0&122&88 (72.1\%) &Cis-regulatory element\\ 
316&6.5&9&3 (33.3\%) &7SK RNA\\ 
1&0.0&2&0 (0.0\%) &Telomerase RNA\\ 
9&0.0&2&1 (50.0\%) &Vault RNA\\ 
892&3.0&3&2 (66.7\%) &Y RNA\\ 
1084&173.5&300&278 (92.7\%) &Transfer RNA\\ 
80&9.5&4&2 (50.0\%) &Transfer RNA pseudogene\\ 
941&3.0&4&2 (50.0\%) &SRP RNA\\ 
607&7.0&22&10 (45.5\%) &Ribosomal RNA\\ 
4&1.0&2&2 (100.0\%) &RNase P/MRP RNA\\ 
\hline
7340&1080.0&1194&865 (72.4\%) &Total\\ 
\hline
\end{tabular}
\caption{}
\label{table:1}
\end{table}

%totalExpressed[865] totalNegControlsExpressed[123] totalGenes[1194] FPR[10.3]

\end{bmcformat}
\end{document}